# Temperature dependence of scintillation properties of SrMoO$_4$

V. B. Mikhailik[1], Yu. Elyashevskyi[2,3], H. Kraus[2], H. J. Kim[4], V. Kapustianyk[3], M. Panasyuk[3]

[1] Diamond Light Source, Harwell Science Campus, Didcot, OX11 0DE, UK
[2] Department of Physics, University of Oxford, Keble Rd., Oxford, OX1 3RH, UK
[3] Scientific-technical and Educational Centre of low Temperature Studies, I. Franko National University of Lviv, 50 Dragomanova Str., 79005, Lviv, Ukraine
[4] Department of Physics of Kyungpook National University, 1370 Sangyeok-dong, Buk-gu, Daegu, 702-701, Republic of Korea

**Abstract**
Studies of the X-ray luminescence and scintillation properties of a SrMoO$_4$ crystal as function of temperature down to T=10 K have been carried out. The luminescence in SrMoO$_4$ is quenched at room temperature, but below T<200 K the crystal exhibits a broad emission band with a maximum at a wavelength of 520 nm. The emission is attributed to the radiative decay of self-trapped excitons and defects acting as traps for the exactions at low temperatures. Such complex character of radiative decay is reflected in the kinetics which contains several components plus a contribution from delayed recombination at low temperatures. The temperature dependence of scintillation light output of SrMoO$_4$ was studied. Comparing with a reference ZnWO$_4$ crystal measured under the same experimental conditions it was found that the light output of SrMoO$_4$ is 15±5%. It is suggested, therefore, that there is scope for optimisation of strontium molybdate for application as scintillator in cryogenic rare event searches.

## 1. Introduction

Molybdate compounds have been and remain interesting materials for study of their scintillation properties. This is driven by a variety of applications as optoelectronic devices such as solid-state lighting [1], [2], acousto-optical elements [3] and lasers hosts [4]. This is due to the combination of appealing physical properties like transparency, ability to accommodate dopants, mechanical and chemical stability as well as availability of technology for production of large single crystals [5]. Scintillation properties of molybdates were studied previously [6] but due to a low light yield the chance for these materials to be used for detection of ionising radiation was negligible. This has changed significantly during the past decade, when development of cryogenic detection technology prompted renewed interest in scintillating molybdates. The main advantage of this technology is the ability to measure the phonon and light response induced by particle interaction in a scintillation target [7] [8]. This allows efficient event discrimination and hence offers a powerful tool for identification of true signals through rejection of radioactive background. The technique has demonstrated its usefulness in low-background physics experiments, searching for rare events such as interactions with Dark Matter particles [9], radioactive decay of long-lived nuclei [10] and neutrinoless double beta decay [11], [12].

As there are many compounds which scintillate at low temperatures it is possible to select the material composition suited for the specific experimental objective. This feature is particularly attractive for experiments searching for neutrinoless double beta decay. Due to a high transition energy (Q=3034 keV) and natural abundance (9.824%) $^{100}$Mo is considered a most promising nucleus for searches for this. Therefore, scintillation crystals that contain molybdenum are actively investigated as potential candidates for such an application.

Scintillation properties of many crystals within the $AMoO_4$ family (A=Pb, Cd, Mg, Ca, Zn) were examined previously over a wide temperature range [8], [13], [14], [15], [16]. Albeit fairly bright, $CdMoO_4$ and $PbMoO_4$ may exhibit enhanced radioactive background due to $^{113}$Cd and $^{210}$Pb. On the other hand, $MgMoO_4$ and $ZnMoO_4$ are rather dull scintillators even at low temperatures[1]. As the discrimination power of cryogenic phonon-scintillation detectors depends on the scintillation light yield, it has been concluded that $CaMoO_4$, exhibiting the highest light yield at cryogenic temperatures [15], is one of the most promising materials. This is one of the reasons for the choice of $CaMoO_4$ for the large scale experiment AMoRE [17]. It has been also acknowledged, however, that this material has one fundamental drawback: the two neutrino double beta decay of $^{48}$Ca with a cut-off energy at 4271 keV will eventually impose a natural limit upon the sensitivity of the experiment searching for neutrinoless double beta decay of $^{100}$Mo [18]. It is possible to reduce this background by two orders of magnitude when using calcium depleted of $^{48}$Ca [19] but this improvement comes at very high price due to the cost of the isotopes used. Therefore, finding an additional scintillating compound with a light yield compatible to that of $CaMoO_4$ remains a well-justified task.

---

[1] Nonetheless, despite of low scintillation light yield, $ZnMoO_4$ is currently actively tested by the LUMINEU collaboration as a prospective material for cryogenic experiments, searching for neutrino-less double beta decay [17].

Earlier experiments carried out at room temperature indicated that although much weaker than $CaMoO_4$, the X-ray luminescence efficiency of $SrMoO_4$ exceeds that of many other molybdates [20]. It is therefore quite sensible to consider strontium molybdate as an additional material. Pirro et al demonstrated the feasibility of $SrMoO_4$ as cryogenic phonon scintillation detector [21] but no further research has been done until now when crystals of good quality were produced by using the Czochralski method and some luminescence and scintillation properties were investigated [22]. Given the interest in this material we decided to carry out additional studies on a $SrMoO_4$ crystal over a wide temperature range.

## 2. Experiment

A $SrMoO_4$ crystal sample of $5\times5\times2$ mm$^3$ was cut from a central part of 15 mm long 10 mm diameter ingot produced by the Czochralski method. For measurements of X-ray luminescence spectra the sample was placed into a closed-cycle He cryostat, equipped with a DE-202A cryocooler (Advanced Research Systems) and Cryocon 32 (Cryogenic Control Systems Inc.) temperature regulator. The emission was excited by a URS-55A X-ray source with a Cu-anticathode tube operating at 55 kV and 10 mA. The luminescence spectra were measured using an automated spectrograph M266 and CCD-camera equipped with a Hamamatsu S7030-1006S sensor, sensitive over a 200 – 1100 nm wavelength range. Thermally stimulated luminescence (TSL) curves where recorded at a constant heating rate of 0.066 K/sec after X-ray irradiation of the crystal at 10 K for 10 minutes.

For measurements of scintillation characteristics the crystal was placed in a helium constant flow cryostat and excited by α-particles from an $^{241}$Am source. The measurements and analysis were carried out using the multi-photon counting technique [23]. The signal, detected by a multi-alkali photomultiplier model 9124A (Electron Tube Enterprises, Ruislip, UK), was digitized by a fast ADC with 5 ns sampling interval. This allows resolving individual photons and recording single photon signals (SPS). The technique allows to measure scintillation processes with decay time constants in the range $10^{-6}$-$10^{-3}$ s as can be observed in many scintillation materials during temperature changes. Furthermore, it allows measuring both decay time and light yield characteristics of scintillators in one experiment. It is well suited for investigation of temperature-dependent scintillator properties as documented in various publications [13], [23].

## 3. Results and discussion

Strontium molybdate has a scheelite-type tetragonal crystal structure with space group *I*41/*a*. The main motif of this structure is an oxyanion complex $MoO_4^{2-}$ surrounded by eight $Sr^{2+}$ cations [24]. Since the bottom of the conduction band and the top of the valence band are formed mainly by molybdenum and oxygen states [25], the oxyanion complex determines the main optical properties of the crystal.

Under X-ray excitation, $SrMoO_4$ exhibits a broad luminescence band in the 400 – 700 nm range that gradually increases in intensity with cooling the crystal (see Fig.1). The emission band has complex character, which is typical for a system where relaxation occurs between the manifold of energy states related to different emission centres. Figure 2 shows the emission spectra as a function of photon energy. The spectrum can be represented as a sum of two Gaussians, indicating that at least two types of emission centres exist in the crystal under study. Such a composite structure of luminescence spectra is a characteristic feature of molybdate crystals [16], [26]. Therefore, as in the case with other molybdates, the high-energy emission band of $SrMoO_4$ is attributed to the radiative decay of self-trapped excitons (STE) localised at the $MoO_4^{2-}$ oxyanion complex while the low energy band is usually interpreted as a emission of STE trapped at defects [27], [28]. Here, this defect-related emission is the dominant contribution to the emission spectra of the $SrMoO_4$ under study here. That can be interpreted as an indication of insufficient quality of the crystal used in this study.

As $SrMoO_4$ exhibits a fairly low temperature of luminescence quenching, measurable scintillation is observed only when the crystal is cooled to temperatures below 200 K. When the crystal is cold enough, it is possible to identify a peak in the pulse height spectra that results from excitation by 5.5 MeV α-particles (see fig.3). The position of this peak is proportional to the amplitude of the scintillation response of the crystal, and thus is used as a measure of scintillation light output at different temperatures. With cooling the crystal the peak shifts towards higher amplitudes, indicating a steady rise of the scintillation light output (see fig. 4). This trend is observed until the temperature reaches 60 K when scintillation light output starts to decrease, as shown in fig.4. At T=12 K the scintillation light output of the $SrMoO_4$ crystal reduces to about 60% of its maximum value. This is consistent with the results obtained in Ref. [22] where light yield was assessed using γ-quanta from a [137]Cs source. It is worthwhile noting that the decrease of scintillation light yield with temperature has been observed in some other molybdates. Such feature can be explained by capture of excited free carriers by shallow traps [14], [29], [28], [30]. This notion is corroborated by correlation between the temperature change of the light output and the TSL response of the crystal under study. The prominent peaks observed between 30 and 50 K in the TSL curve of $SrMoO_4$ (see fig.4, insert) are associated with the thermal release of captured charge carriers. Hence, in this temperature range the probability of trapping increases significantly, reducing the number of charge carriers taking part in the emission. They may contribute to the delayed phosphorescence signal, leading to the appearance of the delayed component in the scintillations. By inspecting the temperature changes in the scintillation decay curves of $SrMoO_4$ displayed in fig.5 the appearance of such a long-lasting component at low temperatures can be clearly identified. The tail of this component may escape detection due to the limited time window used for recording scintillations.

The existence of several types of emission centres with different decay kinetics changing with temperature results in a complex pattern of radiative decay of excitations in $SrMoO_4$. We measure scintillation decay curves in the integral regime thus capturing the entire emission spectrum of the scintillator. Therefore the decay curves reveal contributions of at least three emission processes: i) radiative decay of STE, ii) defect-related emission and iii) delayed radiative recombination of carriers released from traps observed at low temperatures. This complexity imposes certain limitations on the analysis of the decay kinetics. More specifically, the fitting of the scintillation decay curves using a linear combination of exponential functions is merely a mathematical way of representing the experimental results and it is not possible to relate the fitting parameters to the physical quantities that describe the specific emission processes. We used a sum of three exponential functions for fitting to quantify the experimental results obtained from the studies of scintillation decay curves of $SrMoO_4$. The decay time constants as function of temperature are displayed in fig 6. It should be noted that data for the temperature dependence of $\tau_2$ correlate reasonably well with those of [22] while the long decay component $\tau_3$ has not been reported there. This is due to the fact that the multiphoton counting technique is much better suited for investigation of scintillation processes in the millisecond range as opposed to the single photon coincidence technique used in [22]. The decay time constants show a gradual increase with cooling the crystal until the temperature reduces to T=30 K. Below this temperature the carriers captured by shallow traps cannot be anymore released through the thermal activation process. Most likely, the recombination of the carriers of opposite sign occurs through the tunnel process that is consistent with the plateau in the $\tau = f(T)$ dependence below T=30 K.

The results obtained for scintillation light output can be used for the evaluation of the relative light yield of $SrMoO_4$ relative to a reference scintillator. In this case we decided to use a $ZnWO_4$ scintillator as reference which emits in the same spectral range as $SrMoO_4$. Matching emission spectra of two scintillators allows reducing error arising from the uncertainty of calculating the emission-weighted detector efficiency $\varepsilon_\lambda(\lambda)$. The value of $\varepsilon_\lambda(\lambda)$ calculated from the wavelength-dependent emission spectra and the quantum efficiency of the 9124A photomultiplier was found to be 0.16 and 0.14 for $ZnWO_4$ and $SrMoO_4$, respectively. The light collection efficiency of the experimental setup used in this study is determined predominantly by the geometrical factors and as such it can be assumed identical for both, test and reference crystals. Therefore the relative light output can be calculated as a ratio of the measured light outputs of the two crystals corrected for the difference in $\varepsilon_\lambda(\lambda)$. The light output of the reference $ZnWO_4$ measured in the same experimental setup at 10 K is found to be 5.9 times higher than that of $SrMoO_4$. From this we can estimate the relative light yield of $SrMoO_4$ as equal to $(0.14/0.16)\times(1/5.9)\times100\%=15\pm5\%$ of $ZnWO_4$. The error of this evaluation mainly stems from the uncertainty in the measurements of the position of the peak that corresponds to α-events in $SrMoO_4$, shown in fig. 3.

### 4. Conclusion

Scintillation crystals containing molybdenum are important for cryogenic particle physics experiments searching for neutrinoless double beta decay of $^{100}$Mo. Motivated by this we investigated X-ray luminescence, scintillation light output and decay kinetics of a $SrMoO_4$ crystal as function of temperature down to T=10 K. Due to significant thermal quenching, scintillation is observed in the crystal only below T=200 K. The scintillation light output of $SrMoO_4$ increases progressively with decrease of temperature to 60 K and then it reduces by 40% of its maximum value when cooling to 10 K. It is found that at this temperature the light output of strontium molybdate is equal to $15\pm5\%$ of our $ZnWO_4$ reference scintillator. This value is still too low to be considered competitive but the results of this study also show that a significant fraction of excitations created in the crystal is captured by defects or traps. It is expected therefore that scintillation light output can be improved by optimization of the crystal production technique which should yield better quality crystals with fewer defects and traps. It is worth remarking that the low-temperature light output of $CaMoO_4$ crystals initially was only a quarter of the light output of $CaWO_4$ [26], and further development and optimization of calcium molybdate scintillators resulted in equalizing the light output of both scintillators [13].

Taking into account the above fundamental limitations of $CaMoO_4$ in the search for neutrino-less double beta decay of $^{100}$Mo, optimized $SrMoO_4$ can offer a viable alternative. Of course, one should bear in mind the enhanced intrinsic radioactivity of $SrMoO_4$ from the long-lifetime β-emitter $^{90}$Sr. However, the maximum energy of β-decay (2.28 MeV) is below the transition energy in $^{100}$Mo. As such it is no fundamental difficulty and can be mitigated by technical means such as screening of raw materials, their purification, plus improvement of time resolution of the detector to deal with possible pile-up of randomly coinciding events [31]. Therefore, $SrMoO_4$ has the potential of being used in future experiments searching for neutrino-less double beta decay of $^{100}$Mo.

**Acknowledgment**

The study was supported by a grant from the Royal Society (London) ''Cryogenic scintillating bolometers for priority experiments in particle physics'' and the Science & Technology Facilities Council (STFC).

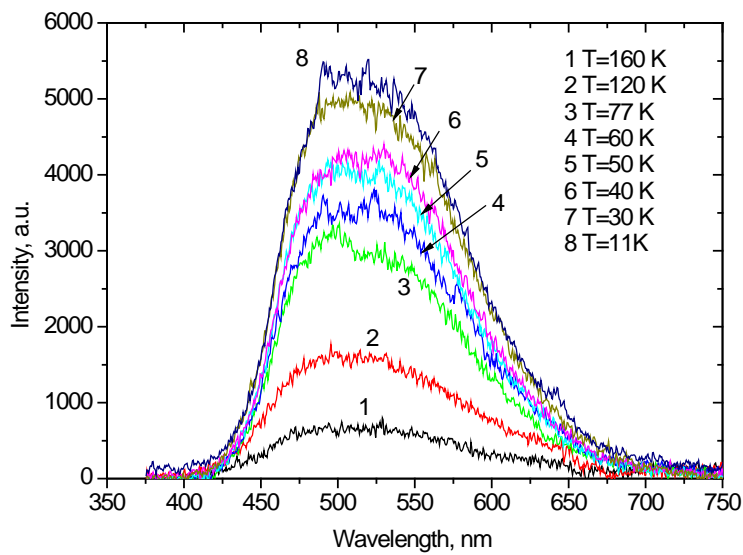

Fig.1: X-ray luminescence spectra of SrMoO$_4$ as function of temperature.

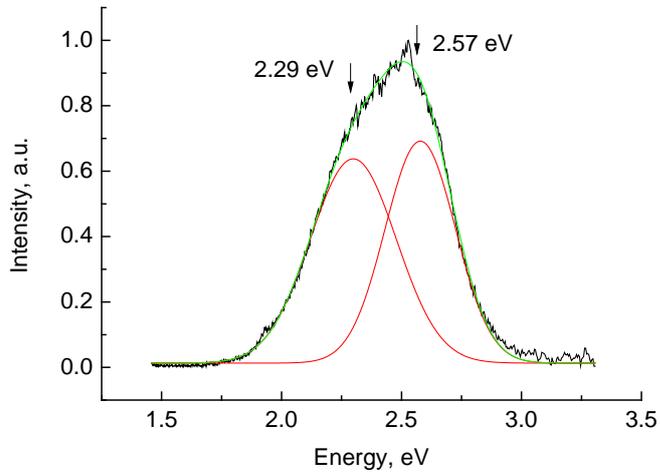

Fig.2: X-ray luminescence spectrum of SrMoO$_4$ measured at T=11 K (black) fitted by two Gaussians (red). The green line presents the sum of the two Gaussians bands.

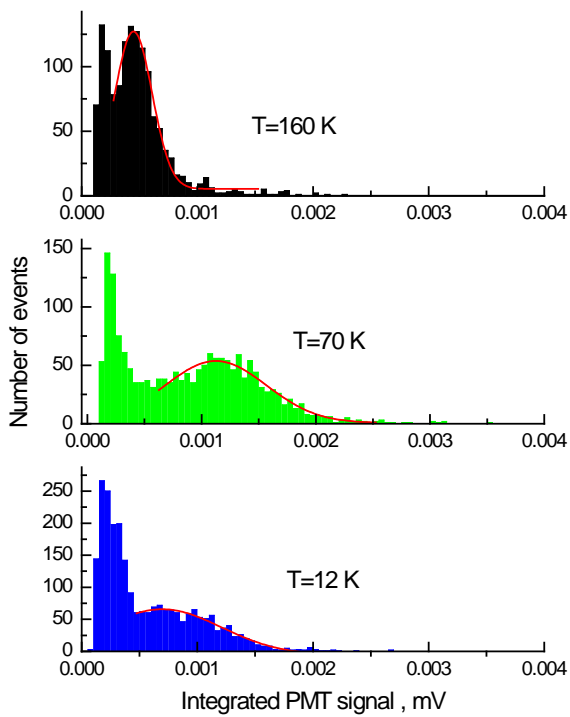

Fig. 3: Pulse height spectra of scintillation response of the SrMoO$_4$ crystal to α-particle excitation from an $^{241}$Am source at different temperatures. Red lines show the Gaussian fitting of the peaks.

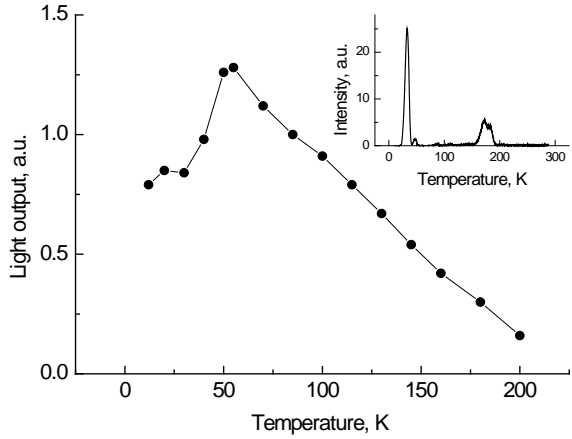

Fig. 4: Light output of SrMoO$_4$ as function of temperature for excitation with α-particles from $^{241}$Am. The insert shows the TSL curve of the crystal monitored after X-ray irradiation of the crystal for 10 minutes.

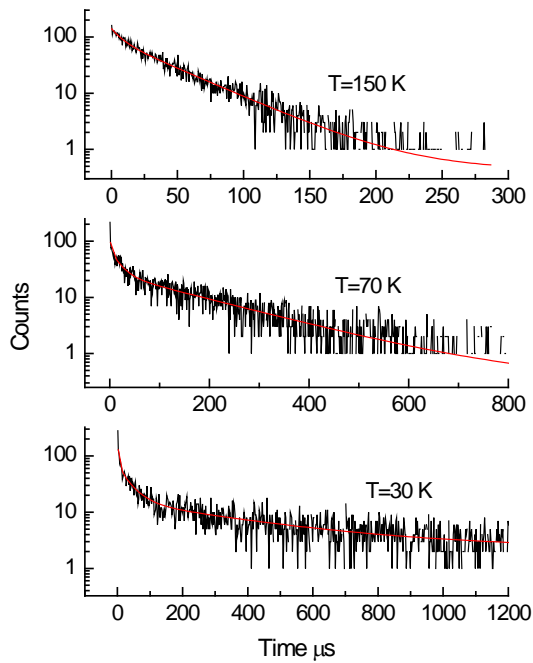

Fig. 5: Decay curves of SrMoO$_4$ measured at T=150, 70 and 30 K for excitation with α-particles from $^{241}$Am. Red lines show the fitting of experimental data using three exponential functions.

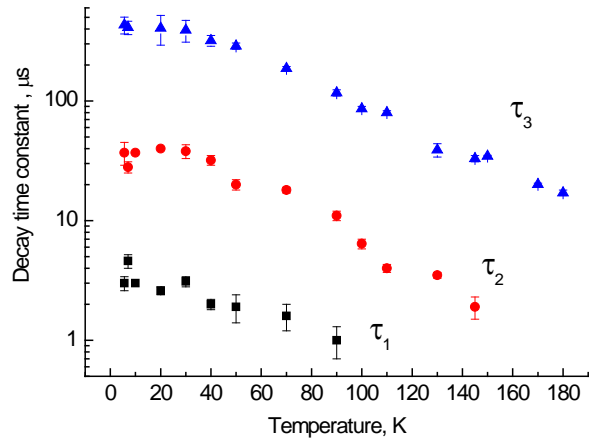

Fig. 6: Temperature dependence of decay time constants of SrMoO$_4$ for excitation with α-particles from $^{241}$Am.